\documentclass[preprint,showpacs,preprintnumbers,eqsecnum,amsmath,amssymb]{revtex4-1}
\usepackage{epsf}
\usepackage{color}
\usepackage{dcolumn}
\usepackage{bm}
\everymath{\displaystyle}

\newcommand{\dE}{I \! \! E}
\newcommand{\dB}{I \! \! B}

\newcommand{\dM}{I \! \! M}

\begin{document} 

\title{Spin in an arbitrary gravitational field}

\author{\firstname{Yuri N.}~\surname{Obukhov}}
\email{obukhov@ibrae.ac.ru}
\affiliation{Theoretical Physics Laboratory, Nuclear Safety Institute,
Russian Academy of Sciences, B. Tulskaya 52, 115191 Moscow, Russia}

\author{\firstname{Alexander J.}~\surname{Silenko}}
\email{alsilenko@mail.ru} \affiliation{Research Institute for
Nuclear Problems, Belarusian State University, Minsk 220030, Belarus,\\
Bogoliubov Laboratory of Theoretical Physics, Joint Institute for Nuclear Research,
Dubna 141980, Russia}

\author{\firstname{Oleg V.}~\surname{Teryaev}}
\email{teryaev@theor.jinr.ru} \affiliation{Bogoliubov Laboratory
of Theoretical Physics, Joint Institute for Nuclear Research,
Dubna 141980, Russia}


\begin {abstract}
We study the quantum mechanics of a Dirac fermion on a curved
spacetime manifold. The metric of the spacetime is completely arbitrary,
allowing for the discussion of all possible inertial and gravitational
field configurations. In this framework, we find the Hermitian Dirac
Hamiltonian for an arbitrary classical external field (including the
gravitational and electromagnetic ones). In order to discuss the physical
content of the quantum-mechanical model, we further apply the Foldy-Wouthuysen
transformation, and derive the quantum equations of motion for the spin and
position operators. We analyse the semiclassical limit of these equations
and compare the results with the dynamics of a classical particle with
spin in the framework of the standard Mathisson-Papapetrou theory and in
the classical canonical theory. The comparison of the quantum mechanical
and classical equations of motion of a spinning particle in an arbitrary
gravitational field shows their complete agreement.
\end{abstract}

\pacs{04.20.Cv; 04.62.+v; 03.65.Sq} \maketitle

\section{Introduction}

Immediately after the notion of spin was introduced in physics and after the
relativistic Dirac theory was formulated, the study of spin dynamics in a curved
spacetime (i.e., in the gravitational field) was initiated. The early efforts were
mainly concerned with the development of mathematical tools and methods appropriate
for the description of the interaction of spinning particles with the gravitational
field \cite{Tetrode,Weyl29,FockIvan,Fock,VDW,Schr,Bargmann,IVDW,Bade,Brill57,VDW60}.
As an interesting by-product, the studies of the spinor analysis in the framework of
the general Lagrange-Noether approach have subsequently resulted in the construction
of the gauge-theoretic models of physical interactions, including also gravity
\cite{Weyl50,Sciama,Kibble,HonnefH,Reader}.

At a later stage, considerable attention was turned to the investigation of
the specific physical effects in the gravitational field predicted for quantum,
semiclassical, and classical relativistic particles with spin
\cite{Oliveira,KO,Lich,Brill66,wong,kann,aud,rudiger,d4,d6,d7,d8,d9,d10,kiefer,goss}.
Various aspects of the dynamics of fermions were studied in the weak gravitational
field, i.e., for the case when the geometry of the spacetime does not significantly
deviate from the flat Minkowski manifold. Another class of problems was the analysis
of trajectories of semiclassical and classical particles in the gravitational field
configurations which arise as the exact solutions of Einstein's equations (such as
the spherically symmetric Schwarzschild metric or the Kerr metric of a rotating
source). The behavior of spin in the strong gravitational fields represents another
interesting subject with the possible applications to the study of the physical
processes in the vicinity of massive astrophysical objects and near black holes.
For the overview of the important mathematical subtleties, the reader can consult
\cite{Benn,esposito,Friedrich,NiRev}, for example.

In this paper, we continue our investigations of the quantum and semiclassical
Dirac fermions using the method of the Foldy-Wouthuysen (FW) transformation. Earlier,
we analyzed the dynamics of spin in weak static and stationary gravitational fields
\cite{PRD,PRD2,Warszawa,OST} and in strong stationary gravitational fields
\cite{ostrong} of massive compact sources. Now we extend our previous results
to the general case of a completely arbitrary gravitational field.

The paper is organized as follows. In Sec.~\ref{Hamiltonian}, we give preliminaries
for the description of the general metric and the coframe, and then derive the
Hermitian Dirac Hamiltonian in an arbitrary curved spacetime.
For completeness, we consider the electrically charged particle interacting also with the
electromagnetic field. In Sec.~\ref{FW1}, we outline the FW technique
and apply this method to derive the FW Hamiltonian together with the corresponding
operator equations of motion. The central result is the derivation of the precession
of spin in an arbitrary gravitational field.
The quantum and semiclassical spin dynamics is compared with
the dynamics of a classical spin in Sec.~\ref{CS}. We use the standard formalism
of Mathisson and Papapetrou, and discuss the Hamiltonian approach. The results
obtained are summarized in Sec.~\ref{final}.

Our notations and conventions are the same as in \cite{PRD2}. In particular, the
world indices of the tensorial objects are denoted by Latin letters $i,j,k,\ldots
= 0,1,2,3$ and the first letters of the Greek alphabet label the tetrad indices,
$\alpha,\beta,\ldots = 0,1,2,3$. Spatial indices of 3-dimensional objects are denoted
by Latin letters from the beginning of the alphabet, $a,b,c,\ldots = 1,2,3$. The 
particular values of tetrad indices are marked by hats.

\section{Dirac particle in a gravitational field}\label{Hamiltonian}

\subsection{General parametrization of the spacetime metric}

Let us recall some basic facts and introduce the notions and objects related to the
description of the motion of a classical spinning particle in a curved manifold. The
massive particle is quite generally characterized by its position in spacetime,
$x^i(\tau)$, where the local spacetime coordinates are considered as functions of
the proper time $\tau$, and by the tensor of spin $S^{\alpha\beta} = -\,S^{\beta\alpha}$.
The analysis of the dynamics of the classical spinning particle is given later in
Sec.~\ref{CS}.

We denote 4-velocity of a particle $U^\alpha = e^\alpha_i dx^i/d\tau$. In view of the
choice of parametrization by the proper time, it is normalized by the standard condition
$g_{\alpha\beta}U^\alpha U^\beta = c^2$ where $g_{\alpha\beta} = {\rm diag}(c^2, -1, -1, -1)$
is the flat Minkowski metric. We use the tetrad $e^\alpha_i$ (or coframe) to describe
the dynamics of spinning particles on a spacetime manifold in arbitrary curvilinear
coordinates. When the spacetime is flat, which means that the gravitational field is
absent, one can choose the global Cartesian coordinates and the holonomic
orthonormal frame that coincides with the natural one, $e^\alpha_i = \delta^\alpha_i$.
The spacetime metric is related to the coframe field in the usual way: $g_{\alpha\beta}
e^\alpha_ie^\beta_i = g_{ij}$.

We use the notation $t$ for the coordinate time and $x^a~ (a=1,2,3)$ denote spatial
local coordinates. There are many different ways to represent a general spacetime
metric. A convenient parametrization of the spacetime metric was proposed by De Witt
\cite{Dewitt} in the context of the canonical formulation of the quantum gravity theory.
In a slightly different disguise, the general form of the line element of an arbitrary
gravitational field reads
\begin{equation}\label{LT}
ds^2 = V^2c^2dt^2 - \delta_{\widehat{a}\widehat{b}}W^{\widehat
a}{}_c W^{\widehat b}{}_d\,(dx^c - K^ccdt)\,(dx^d - K^dcdt).
\end{equation}
This parametrization involves more functions than the actual number of the metric
components. Indeed, the total number of the functions $V(t,x^c)$, $K^a(t,x^c)$, and
$W^{\widehat a}{}_b(t,x^c)$ is $1+3+9=13$ which is greater than 10. However, we have to
take into account that the line element (\ref{LT}) is invariant under redefinitions
$W^{\widehat a}{}_b\longrightarrow L^{\widehat a}{}_{\widehat c}W^{\widehat c}{}_b$ using arbitrary
local rotations $L^{\widehat a}{}_{\widehat c}(t,x)\in SO(3)$. Subtracting the 3 rotation
degrees of freedom, we recover exactly 10 independent variables that describe the
general metric of the spacetime.

\subsection{Dirac equation}

One needs the orthonormal frames to discuss the spinor field and to
formulate the Dirac equation. From the mathematical point of view, the tetrad is
necessary to ``attach'' a spinor space at every point of the spacetime manifold.
Tetrads (coframes) are naturally defined up to a local Lorentz transformations,
and one usually fixes this freedom by choosing a gauge. We discussed the choice
of the tetrad gauge in \cite{OST} and have demonstrated that a physically preferable
option is the Schwinger gauge \cite{Schwinger,dirac}, namely the  condition
$e_a^{\,\widehat{0}} =0, a=1,2,3$. Accordingly, for the general metric (\ref{LT}) we will
work with the tetrad
\begin{equation}\label{coframe}
e_i^{\,\widehat{0}} = V\,\delta^{\,0}_i,\qquad e_i^{\widehat{a}} =
W^{\widehat a}{}_b\left(\delta^b_i - cK^b\,\delta^{\,0}_i\right),\qquad a=1,2,3.
\end{equation}
The inverse tetrad, such that $e^i_\alpha e^\alpha_j = \delta^i_j$,
\begin{equation}\label{frame}
e_{\,\widehat{0}}^i = {\frac 1V}\left(\delta_{\,0}^i +
\delta_{\,a}^icK^a\right), \qquad e^i_{\widehat{a}} =
\delta_{\,b}^iW^b{}_{\widehat a},\qquad a=1,2,3,
\end{equation}
also satisfies the similar Schwinger condition, $e^0_{\,\widehat{a}} = 0$. Here we
introduced the inverse $3\times 3$ matrix, $W^a{}_{\widehat c}W^{\widehat c}{}_b = \delta_b^a$.

The following observation will be useful for the subsequent computations. A classical
massive particle moves along a world line $x^i(\tau), i=0,1,2,3$, parametrized by
the proper time $\tau$. Its 4-velocity is defined as usual by the derivatives
$U^i = dx/d\tau$. With respect to a given orthonormal frame, the velocity has the
components $U^\alpha = e^\alpha_iU^i, \alpha = 0,1,2,3$. It is convenient to describe
the 4-velocity by its 3 spatial components $v^{\widehat a}, a=1,2,3$, in an anholonomic
frame. Then $U^\alpha = (\gamma, \gamma v^{\widehat a})$, with the Lorentz factor
$\gamma^{-1} = \sqrt{1 - v^2/c^2}$, and, consequently,
\begin{eqnarray}\label{U0}
U^0 &=& {\frac {dt}{d\tau}} = e^0_\alpha U^\alpha = {\frac \gamma V},\\
U^a &=& {\frac {dx^a}{d\tau}} = e^a_\alpha U^\alpha = {\frac \gamma V}\left(
cK^a + VW^a{}_{\widehat b}\,v^{\widehat b}\right).\label{Ua}
\end{eqnarray}
We used (\ref{frame}) here. Dividing (\ref{Ua}) by (\ref{U0}) and denoting
$${\cal F}^a{}_b = VW^a{}_{\widehat b},$$ we find for the velocity with respect to the
coordinate time
\begin{equation}
{\frac {dx^a}{dt}} = {\cal F}^a{}_b\,v^b + cK^a.\label{dtv}
\end{equation}

The Dirac equation in a curved spacetime reads
\begin{equation}
(i\hbar\gamma^\alpha D_\alpha - mc)\Psi=0,\qquad \alpha=0,1,2,3.
\label{Dirac0}
\end{equation}
This equation is invariant under the general transformations of the spacetime coordinates
(under diffeomorphism), and is covariant under the local Lorentz transformations. Recall
that the Dirac matrices $\gamma^\alpha$ are defined in local Lorentz (tetrad) frames and
they have constant components. The spinor covariant derivatives are consistently defined
in the gauge-theoretic approach \cite{Sciama,Kibble,HonnefH,Reader} as
\begin{equation}
D_\alpha = e_\alpha^i D_i,\qquad D_i = \partial _i + {\frac {iq}{\hbar}}
\,A_i + {\frac i4}\sigma^{\alpha\beta}\Gamma_{i\,\alpha\beta}.\label{eqin2}
\end{equation}
Here the Lorentz connection is $\Gamma_i{}^{\alpha\beta} = - \Gamma_i{}^{\beta\alpha}$, and
$\sigma^{\alpha\beta} = {\frac i2}\left(\gamma^\alpha \gamma^\beta - \gamma^\beta\gamma^\alpha
\right)$ are the generators of the local Lorentz transformations of the spinor field.
For completeness, we assumed that the Dirac particle is charged  and the electric charge
$q$ describes its coupling to the 4-potential of the electromagnetic field $A_i$.
Note that the canonical dimension of the electromagnetic field strength 2-form $F = dA$
and of the electromagnetic 1-form $A = A_idx^i$ is $[F] = [A] = [\hbar/q]$, see \cite{birk}.
The gravitational and inertial effects (which are deeply related to each other in the
framework of the gauge-theoretic approach) are encoded in coframe and connection
in (\ref{Dirac0}),(\ref{eqin2}); for the relevant discussion see Refs. \cite{HN,Ob1,Ob2}
and references therein.

Using the parametrization of the general metric (\ref{LT}) with the tetrad
(\ref{coframe}) in the Schwinger gauge, we find explicitly the components of connection
\begin{eqnarray}
\Gamma_{i\,\widehat{a}\widehat{0}} &=& {\frac {c^2}V}\,W^b{}_{\widehat{a}}
\,\partial_bV\,e_i{}^{\widehat{0}} - {\frac cV}\,{\cal Q}_{(\widehat{a}
\widehat{b})}\,e_i{}^{\widehat{b}},\label{connection1}\\
\Gamma_{i\,\widehat{a}\widehat{b}} &=& {\frac cV}\,{\cal Q}_{[\widehat{a}
\widehat{b}]}\,e_i{}^{\widehat{0}} + \left({\cal C}_{\widehat{a}\widehat{b}
\widehat{c}} + {\cal C}_{\widehat{a}\widehat{c}\widehat{b}} + {\cal C}_{\widehat{c}
\widehat{b}\widehat{a}}\right) e_i{}^{\widehat{c}}.\label{connection2}
\end{eqnarray}
Here we introduced the two useful objects:
\begin{eqnarray}
{\cal Q}_{\widehat{a}\widehat{b}} &=& g_{\widehat{a}\widehat{c}}W^d{}_{\widehat{b}}
\left({\frac 1c}\dot{W}^{\widehat c}{}_d + K^e\partial_e{W}^{\widehat c}{}_d
+ {W}^{\widehat c}{}_e\partial_dK^e\right),\label{Qab}\\
{\cal C}_{\widehat{a}\widehat{b}}{}^{\widehat{c}} &=& W^d{}_{\widehat{a}}
W^e{}_{\widehat{b}}\,\partial_{[d}W^{\widehat{c}}{}_{e]},\qquad {\cal
C}_{\widehat{a} \widehat{b}\widehat{c}} = g_{\widehat{c}\widehat{d}}
\,{\cal C}_{\widehat{a}\widehat{b}}{}^{\widehat{d}}.\label{Cabc}
\end{eqnarray}
As usual, the dot $\dot{\,}$ denotes the partial derivative with respect
to the coordinate time $t$. One can immediately recognize that ${\cal
C}_{\widehat{a}\widehat{b}}{}^{\widehat{c}} =- {\cal C}_{\widehat{b}\widehat{a}}{}^{\widehat{c}}$
is the anholonomity object for the spatial triad ${W}^{\widehat a}{}_b$.
The indices (that all run from 1 to 3) are raised and lowered with
the help of the spatial part of the flat Minkowski metric,
$g_{\widehat{a}\widehat{b}} = -\,\delta_{ab} = {\rm diag}(-1, -1, -1)$.

One can derive the Dirac equation from the action integral
$I = \int\,d^4x\,\sqrt{-g}\,L$, 
with the Lagrangian (recall for the conjugate spinor $\overline{\Psi} := \Psi^\dagger
\gamma^{\widehat{0}}$)
\begin{equation}\label{LD}
L = {\frac {i\hbar}{2}}\left(\overline{\Psi}\gamma^\alpha
D_\alpha\Psi - D_\alpha\overline{\Psi}\gamma^\alpha\Psi\right) -
mc\,\overline{\Psi}\Psi.
\end{equation}
A direct check shows that, with (\ref{eqin2})-(\ref{Cabc}) inserted, the Schr\"odinger
form of the Dirac equation derived from this action involves a non-Hermitian
Hamiltonian. However, this problem is fixed if we introduce a new wave function by
\begin{equation}
\psi = \left(\sqrt{-g}e_{\widehat{0}}^0\right)^{\frac
12}\,\Psi.\label{newpsi}
\end{equation}
Such a non-unitary transformation also appears in the framework of
the pseudo-Hermitian quantum mechanics \cite{GorNezn,GorNeznnew} (cf. \cite{lec}).

Variation of the action with respect to the
{\it rescaled} wave function $\psi$ yields the Dirac equation in Schr\"odinger
form $i\hbar\frac{\partial \psi} {\partial t}= {\cal H}\psi$. The
corresponding {\it Hermitian} Hamiltonian reads
\begin{eqnarray}
{\cal H} &=& \beta mc^2V + q\Phi + {\frac c 2}\left(\pi_b\,{\cal F}^b{}_a
\alpha^a + \alpha^a{\cal F}^b{}_a\pi_b\right)\nonumber\\
\label{Hamilton1} && +\,{\frac c2}\left(\bm{K}\cdot\bm{\pi} +
\bm{\pi}\cdot\bm{K}\right) + {\frac {\hbar c}4}\left(\bm{\Xi}\cdot\bm{\Sigma}
- \Upsilon\gamma_5 \right).
\end{eqnarray}
Here ${\bm K} = \{K^a\}$, and the kinetic momentum operator ${\bm\pi} = \{\pi_a\}$
with $\pi_a = -\,i\hbar\partial_a + qA_a = p_a + qA_a$ accounts for the interaction
with the electromagnetic field $A_i = (\Phi, A_a)$. To remind the notation: $\beta
= \gamma^{\hat 0}, {\bm\alpha} = \{\alpha^a\}, {\bm\Sigma} = \{\Sigma^a\}$, where the
3-vector-valued Dirac matrices have their usual form; namely, $\alpha^a = \gamma^{\hat 0}
\gamma^a$ ($a,b,c,\dots = 1,2,3$) and $\Sigma^1 = i\gamma^{\hat 2}\gamma^{\hat 3},
\Sigma^2 = i\gamma^{\hat 3}\gamma^{\hat 1}, \Sigma^3 = i\gamma^{\hat 1}\gamma^{\hat 2}$.
We also introduced a pseudoscalar $\Upsilon$ and a 3-vector ${\bm\Xi} = \{\Xi_a\}$ by
\begin{equation}
\Upsilon = V\epsilon^{\widehat{a}\widehat{b}\widehat{c}}\Gamma_{\widehat{a}
\widehat{b}\widehat{c}} = - V\epsilon^{\widehat{a}\widehat{b}\widehat{c}}
{\cal C}_{\widehat{a}\widehat{b}\widehat{c}},\qquad \Xi_{\widehat{a}} =
{\frac Vc}\,\epsilon_{\widehat{a}\widehat{b}\widehat{c}}\,\Gamma_{\widehat{0}}
{}^{\widehat{b}\widehat{c}} = \epsilon_{\widehat{a}\widehat{b}\widehat{c}}
\,{\cal Q}^{\widehat{b}\widehat{c}}.\label{AB}
\end{equation}
Note that we have fixed a number of small points with the signs and numeric factors,
and one should be careful when comparing formulas above with the earlier results
in \cite{ostrong}. For the static and stationary rotating configurations, the
pseudoscalar invariant vanishes ($\epsilon^{\widehat{a}\widehat{b}\widehat{c}} {\cal
C}_{\widehat{a}\widehat{b}\widehat{c}} = 0$), and thus the corresponding term was absent in
the special cases considered earlier \cite{OST,ostrong}. But in general this term
contributes to the Dirac Hamiltonian.

It is worthwhile to mention that the recent discussion \cite{GM} of the
Dirac fermions in an arbitrary gravitational field is very
different in that the non-Hermitian Hamiltonian is used in that
work, in deep contrast to the explicitly Hermitian one
(\ref{Hamilton1}).

\section{The Foldy-Wouthuysen transformation}\label{FW1}

In the previous section, we described the dynamics of the fermion in
Dirac representation. The physical contents of the theory is however
revealed in the Foldy-Wouthuysen representation. We will now construct
the FW \cite{FW} Hamiltonian for the fermion moving in an arbitrary
gravitational field described by the general metric (\ref{LT}). We
start with the exact Dirac Hamiltonian (\ref{Hamilton1}) and apply
the method developed in \cite{PRA}.

Just like before in our earlier work \cite{OST,ostrong}, we do not make any
assumptions and/or approximations for the functions $V, W^{\widehat a}{}_b, K^a$.
The Planck constant $\hbar$ will be treated as the only small parameter.
In accordance with this strategy, we retain in the FW Hamiltonian all the
terms of the zero and first orders in $\hbar$. The leading nonvanishing terms
of order $\hbar^2$ have been calculated in both nonrelativistic and weak
field approximations in our previous works \cite{PRD,OST,ostrong} for the more
special cases. These terms describe the gravitational contact (Darwin) interaction.

\subsection{General preliminaries}

A generic Hamiltonian can be decomposed into operators that commute and
anticommute with $\beta$:
\begin{equation}\label{Vvedenieq3}
{\cal H}=\beta {\cal M}+{\cal E}+{\cal O},\quad\beta{\cal M}={\cal M}\beta,
\quad\beta{\cal E}={\cal E}\beta,\quad\beta{\cal O}=-{\cal O}\beta.
\end{equation}
Here, the operators ${\cal M},{\cal E}$ are even, and ${\cal O}$ is odd.

Foldy-Wouthuysen representation is obtained by means of the unitary
transformation
\begin{equation}
\psi_{FW} = U\psi,\qquad {\cal H}_{FW} = U{\cal H}U^{-1}
- i\hbar U\partial_tU^{-1}.\label{FWT0}
\end{equation}
In arbitrary strong external fields, the following transformation operator
can be used \cite{PRA}:
\begin{equation}
U = \frac{\beta\epsilon+\beta {\cal M}-{\cal
O}}{\sqrt{(\beta\epsilon+\beta {\cal M}-{\cal O})^2}}\,\beta,\qquad
U^{-1}=\beta\,\frac{\beta\epsilon+\beta{\cal M}-{\cal
O}}{\sqrt{(\beta\epsilon+\beta{\cal M}-{\cal O})^2}}.\label{4eq18}
\end{equation}
Here  $\epsilon=\sqrt{{\cal M}^2+{\cal O}^2}$, and $U^{-1} = U^\dagger$ if
${\cal H}={\cal H}^\dagger$. Applying (\ref{FWT0}), we obtain the explicit
transformed Hamiltonian
\begin{eqnarray}
{\cal H}' &=& \beta\epsilon+{\cal E}+ \frac{1}{2T}\Biggl(\left[T,\left[
T,(\beta\epsilon+{\cal Z})\right]\right] +\beta\left[{\cal O},[{\cal O},
{\cal M}]\right]\nonumber\\
&& - \left[{\cal O},\left[{\cal O},{\cal Z}\right]\right] - \left[(\epsilon
+{\cal M}),\left[(\epsilon+{\cal M}),{\cal Z}\right]\right] - \left[(\epsilon
+{\cal M}),\left[{\cal M},{\cal O}\right]\right]\nonumber\\
&& -\beta \left\{{\cal O},\left[(\epsilon+{\cal M}),{\cal Z}\right]\right\}
+\beta \left\{(\epsilon+{\cal M}),\left[{\cal O},{\cal Z}\right]\right\}
\Biggr)\frac{1}{T},\label{eq28N}
\end{eqnarray}
where ${\cal Z}={\cal E}-i\hbar\frac{\partial}{\partial t}$ and
$T=\sqrt{(\beta\epsilon+\beta{\cal M}-{\cal O})^2}$. The square and curly
brackets denote the commutator $[A,B] = AB - BA$ and
the anticommutator $\{A,B\} = AB + BA$, respectively.

The Hamiltonian (\ref{eq28N}) still contains odd terms proportional to $\hbar$.
We can write it as follows:
\begin{equation}
{\cal H}' = \beta\epsilon+{\cal E}'+{\cal O}',\quad\beta{\cal E}'={\cal
E}'\beta, \quad\beta{\cal O}'=-{\cal O}'\beta,
\label{4eq27}
\end{equation}
where $\epsilon=\sqrt{{\cal M}^2+{\cal O}^2}$. The even and odd parts are
determined by
\begin{equation}
{\cal E}'=\frac12\left({\cal H}'+\beta{\cal H}'\beta\right)-\beta\epsilon,\qquad
{\cal O}'=\frac12\left({\cal H}'-\beta{\cal H}'\beta\right).\label{EO}
\end{equation}
Additional unitary transformation removes the odd part, so that
the final approximate Hamiltonian reads \cite{PRA}
\begin{equation}
{\cal H}_{FW}=\beta\epsilon+{\cal E}'+\frac14\beta\left\{{\cal
O}'^2,\frac{1}{\epsilon}\right\}.\label{eqf}
\end{equation}

For the case under consideration, we have explicitly
\begin{eqnarray}
{\cal M} &=& mc^2V,\\
{\cal E} &=& q\Phi + {\frac c2}\left(\bm{K}\cdot\bm{\pi} +
\bm{\pi}\cdot\bm{K}\right) + {\frac {\hbar c}4}\,\bm{\Xi}\cdot\bm{\Sigma},\\
{\cal O} &=& {\frac c 2}\left(\pi_b\,{\cal F}^b{}_a \alpha^a + \alpha^a
{\cal F}^b{}_a\pi_b\right) - {\frac {\hbar c}4}\,\Upsilon\gamma_5.
\end{eqnarray}

\subsection{Foldy-Wouthuysen Hamiltonian and quantum dynamics}

We now limit ourselves to the case when the electromagnetic field is switched off.
The computations along the lines described in the previous subsection are
straightforward, and after a lengthy algebra we obtain the Foldy-Wouthuysen
Hamiltonian in the following form
\begin{equation}
{\cal H}_{FW}={\cal H}_{FW}^{(1)}+{\cal H}_{FW}^{(2)}.\label{eqFW}
\end{equation}
Here the two terms read, respectively, 
\begin{widetext}
\begin{eqnarray}
{\cal H}_{FW}^{(1)} &=& \beta\epsilon' + \frac{\hbar c^2}{16}\left\{
\frac{1}{\epsilon'},\left(2\epsilon^{cae}\Pi_e \{p_b,{\cal F}^d{}_c\partial_d
{\cal F}^b{}_a\}+\Pi^a\{p_b,{\cal F}^b{}_a\Upsilon\}\right)\right\}\nonumber\\
&& +\,\frac{\hbar mc^4}{4}\epsilon^{cae}\Pi_e\left\{\frac{1}{{\cal T}},\left\{
p_d,{\cal F}^d{}_c{\cal F}^b{}_a\partial_bV\right\}\right\},\label{eq7}
\end{eqnarray}
\begin{eqnarray}
{\cal H}_{FW}^{(2)} &=& \frac c2\left(K^a p_a + p_a K^a\right)
+ {\frac {\hbar c}4}\,\Sigma_a\Xi^a\nonumber\\
&& +\,\frac{\hbar c^2}{16}\Biggl\{\frac{1}{{\cal
T}},\biggl\{\Sigma_a \{p_e,{\cal
F}^e{}_b\},\Bigl\{p_f,\bigl[\epsilon^{abc}({\frac 1c} \dot{\cal
F}^f{}_c - {\cal F}^d{}_c\partial_dK^f + K^d\partial_d{\cal
F}^f{}_c)\nonumber\\
&& -\,{\frac 12}{\cal F}^f{}_d\left(\delta^{db}\Xi^a -
\delta^{da}\Xi^b\right) \bigr]\Bigr\}\biggr\}\Biggr\},\label{eq7K}
\end{eqnarray}
\end{widetext}
\begin{equation}\label{eqa}
\epsilon'=\sqrt{m^2c^4V^2+\frac{c^2}{4}\delta^{ac}\{p_b,{\cal F}^b{}_a\}
\{p_d,{\cal F}^d{}_c\}},\qquad {\cal T}=2{\epsilon'}^2 + \{\epsilon',mc^2V\}.
\end{equation}

Let us derive the equation of motion of spin. As usual, we introduce the
polarization operator $\bm\Pi=\beta\bm\Sigma$, and the corresponding dynamical
equation is obtained from its commutator with the FW Hamiltonian. 
The computation is straightforward and we find
\begin{equation}
\frac{d\bm \Pi}{dt}=\frac{i}{\hbar}[{\cal H}_{FW},\bm \Pi]=\bm\Omega_{(1)}
\times\bm \Sigma+\bm\Omega_{(2)}\times\bm\Pi.\label{spinmeq}
\end{equation}
Here the 3-vectors $\bm\Omega_{(1)}$ and $\bm\Omega_{(2)}$ are the operators of
the angular velocity of spin precessing in the exterior classical gravitational
field. Their components read explicitly as follows:
\begin{eqnarray}
\Omega^a_{(1)} &=& \frac{mc^4}{2}\left\{{\frac 1{\cal T}}, \{p_e,
\epsilon^{abc}{\cal F}^e{}_b{\cal F}^d{}_c\partial_d\,V\}\right\}\nonumber\\
&& +\,\frac{c^2}{8}\left\{
\frac{1}{\epsilon'},\{p_e,(2\epsilon^{abc} {\cal
F}^d{}_b\partial_d{\cal F}^e{}_c + \delta^{ab}{\cal
F}^e{}_b\,\Upsilon)\} \right\},\label{eqol}
\end{eqnarray}
and
\begin{eqnarray}
\Omega^a_{(2)} &=& \frac{\hbar c^2}{8}\Biggl\{\frac{1}{{\cal
T}},\biggl\{ \{p_e,{\cal
F}^e{}_b\},\Bigl\{p_f,\bigl[\epsilon^{abc}({\frac 1c} \dot{\cal
F}^f{}_c - {\cal F}^d{}_c\partial_dK^f + K^d\partial_d{\cal
F}^f{}_c
)\nonumber\\
&& -\,{\frac 12}{\cal F}^f{}_d\left(\delta^{db}\Xi^a -
\delta^{da}\Xi^b\right) \bigr]\Bigr\}\biggr\}\Biggr\} + {\frac
c2}\,\Xi^a. \label{finalOmega}
\end{eqnarray}

One may notice that the two different matrices, $\bm\Sigma$ and $\bm\Pi$, appear 
on the right-hand side of Eq. (\ref{spinmeq}). This is explained by the fact that 
the vector $\bm\Omega_{(1)}$ contains odd number of components of the momentum operator,
whereas the vector $\bm\Omega_{(2)}$ contains even number of $p_a$. Actually,
both $\bm\Omega_{(1)}$ and $\bm\Omega_{(2)}$ depend only on the combination
${\cal F}^b{}_a p_b$. However, the velocity operator is proportional to an
additional $\beta$ factor and is equal to $v_a = \beta c^2{\cal F}^b{}_a
p_b/\epsilon'$, as we demonstrate below, see (\ref{Fpv}). As a result, the
operator $\bm\Omega_{(1)}$ also acquires an additional $\beta$ factor \cite{OST},
when it is rewritten in terms of the velocity operator $\bm v$. Note also that 
in the FW representation only upper part of $\beta$ proportional the unit matrix 
is relevant. Therefore, the appearance of $\beta$ does not lead to any physical 
effects at least until antiparticles are considered (which would require 
special investigations).

We now use the general results above to obtain the corresponding semiclassical
expressions by evaluating all anticommutators and neglecting the powers of $\hbar$
higher than 1. Then equations (\ref{spinmeq})-(\ref{finalOmega}) yield the
following explicit semiclassical equations describing the motion of the average spin 
(as usual, vector product is defined by $\{\bm A\times\bm B\}_a=\epsilon_{abc}A^bB^c$):
\begin{eqnarray}
{\frac {d{\bm s}}{dt}} &=& \bm \Omega\times{\bm s}
= (\bm \Omega_{(1)}+\bm \Omega_{(2)})\times{\bm s},\label{ds}\\
\Omega^a_{(1)} &=& {\frac {c^2}{\epsilon'}}{\cal F}^d{}_c p_d\left({\frac 12}
{\Upsilon}\delta^{ac} - \epsilon^{aef}V{\cal C}_{ef}{}^c + {\frac {\epsilon'}
{\epsilon' + mc^2V}}\epsilon^{abc}W^e{}_{\widehat{b}}\,\partial_eV\right),
\label{FO}\\
\Omega^a_{(2)} &=& {\frac c2}\,\Xi^a - {\frac {c^3}{\epsilon'(\epsilon'+mc^2V)}}
\,\epsilon^{abc}Q_{(bd)}\delta^{dn}{\cal F}^k{}_np_k{\cal F}^l{}_cp_l,
\label{finalOmegase}
\end{eqnarray}
where, in the semiclassical limit,
\begin{equation}
\epsilon' = \sqrt{m^2c^4V^2 + c^2\delta^{cd}{\cal F}^a{}_c\,{\cal F}^b{}_d
\,p_a\,p_b\,}.\label{eQ}
\end{equation}

We can substitute the results obtained into the FW Hamiltonian (\ref{eqFW})
and recast the latter in a compact and transparent form in terms of the
precession angular velocities $\bm\Omega_{(1)},\bm\Omega_{(2)}$:
\begin{eqnarray}
{\cal H}_{FW} = \beta\epsilon'+\frac c2\left(\bm K\!\cdot\!\bm p
+ \bm p\cdot\!\bm K\right) +\frac\hbar2\bm\Pi\cdot\bm\Omega_{(1)}+
\frac\hbar2\bm\Sigma\cdot\bm\Omega_{(2)}.\label{Hamlt}
\end{eqnarray}

We can use (\ref{Hamlt}) to derive the velocity operator in the semiclassical
approximation:
\begin{equation}
{\frac {dx^a}{dt}} = \frac{i}{\hbar}[{\cal H}_{FW},x^a] = \beta\,{\frac
{\partial\epsilon'}{\partial p_a}} + cK^a
= \beta\,{\frac {c^2}{\epsilon'}}\,{\cal F}^a{}_b\delta^{bc}{\cal F}^d{}_c
p_d + cK^a.\label{velocity}
\end{equation}
Comparing this with the relation between the holonomic and anholonomic
components of the velocity, (\ref{dtv}),
we find the velocity operator in the Schwinger frame (\ref{coframe}):
\begin{equation}
\beta\,{\frac {c^2}{\epsilon'}}\,{\cal F}^b{}_ap_b = v_a.\label{Fpv}
\end{equation}
This immediately yields $\delta^{cd}{\cal F}^a{}_c\,{\cal F}^b{}_d p_ap_b =
(\epsilon')^2v^2/c^2$. Using this in (\ref{eQ}), we have $(\epsilon')^2 =
m^2c^4V^2 + (\epsilon')^2v^2/c^2$, and thus we find
\begin{equation}
\epsilon' = \gamma\,mc^2\,V.\label{eV}
\end{equation}
Equations (\ref{Fpv}) and (\ref{eV}) are crucial for establishing the full
agreement of the quantum and classical dynamics of spin. In particular, using
(\ref{Fpv}) and (\ref{eV}), we find
\begin{equation}
{\frac {\epsilon'}{\epsilon' + mc^2V}} = {\frac \gamma {1 +
\gamma}},\qquad {\frac {c^3}{\epsilon'(\epsilon'+mc^2V)}}\,{\cal
F}^b{}_ap_b{\cal F}^d{}_cp_d = {\frac \gamma {1 + \gamma}}\,{\frac
{v_av_c}{c}}.\label{factors}
\end{equation}

\subsection{Quantum-mechanical equations of particle dynamics} \label{forces}

We now turn to the analysis of the motion of the quantum particle in the
gravitational field. The dynamics of spin is described in an anholonomic frame.
For consistency, we will use an anholonomic frame description for the particle
dynamics, too. The Schwinger gauge with $e^0_{\widehat a} = 0$ simplifies the
equation for the force operator which is given by
\begin{eqnarray}
F_{\widehat a} &=& {\frac {dp_{\widehat a}}{dt}} = {\frac 12} {\frac d{dt}}
\left\{e^b_{\widehat a},p_b\right\} =  {\frac 12}\left\{{\frac
{dW^b{}_{\widehat a}}{dt}},p_b\right\} + {\frac 12}\left\{W^b{}_{\widehat a},
{\frac {dp_b}{dt}}\right\} \nonumber\\
&=& {\frac 12} \left\{\dot{W}^b{}_{\widehat a},p_b\right\} +  {\frac i{2\hbar}}
\left\{[{\cal H}_{FW},W^b{}_{\widehat a}],p_b\right\} - {\frac 12}\left\{
W^b{}_{\widehat a},\partial_b{\cal H}_{FW}\right\}.\label{forcefactoan}
\end{eqnarray}
Here as before the partial derivative with respect to the coordinate time is denoted
by the dot, in particular, $\dot{W}^b{}_{\widehat a} := \partial_t{W}^b{}_{\widehat a}$.

The explicit expression for the force operator reads
\begin{eqnarray}
F_{\widehat a} &=& {\frac 12} \left\{\dot{W}^b{}_{\widehat a},p_b\right\} +
 {\frac 14} \left\{p_b,\left\{\frac{\partial {\cal H}_{FW}}{\partial p_c},
\partial_cW^b{}_{\widehat a}\right\}\right\} - {\frac 12}\left\{W^b{}_{\widehat a},
\partial_b{\cal H}_{FW}\right\},\label{forcefacto}\\
\frac{\partial {\cal H}_{FW}}{\partial p_c} &=& \beta\frac{c^2}{4}\delta^{ad}
\left\{\frac{1}{\epsilon'}, \left\{p_b, {\cal F}^b{}_a{\cal F}^c{}_d\right\}
\right\} + cK^c + \frac\hbar2\mathfrak{T}^c,\label{dpH}
\end{eqnarray}
where we introduced the following compact notation
\begin{equation}
\mathfrak{T}^c=\frac{\partial {\cal U}}{\partial p_c},\qquad {\cal U}
:= \bm\Pi\cdot\bm\Omega_{(1)} + \bm\Sigma\cdot\bm\Omega_{(2)}.\label{mathfrakT}
\end{equation}
Corrections due to the noncommutativity of operators are of order of $\hbar^2$
and can be neglected in (\ref{forcefacto}). Let us split the total force operator
into the terms of the zeroth and first orders in the Planck constant:
\begin{equation}
F_{\widehat a} = F_{\widehat a}^{(0)}+ F_{\widehat a}^{(1)}.\label{FFF}
\end{equation}
The zeroth order terms read as follows
\begin{eqnarray}
 F_{\widehat a}^{(0)} &=& {\frac 12} \left\{\dot{W}^b{}_{\widehat a},p_b\right\}
- {\frac 12}\left\{W^b{}_{\widehat a},\partial_b\left [\beta\epsilon' +
\frac c2\left(K^a p_a + p_a K^a\right) \right]\right\}\nonumber\\
&& +\,{\frac 14} \left\{p_b,\left\{\left(\beta\frac{c^2}{4}\delta^{ad}
\left\{\frac{1}{\epsilon'},\left\{p_b, {\cal F}^b{}_a{\cal F}^c{}_d\right\}
\right\}+cK^c\right),\partial_cW^b{}_{\widehat a}\right\}\right\}.\label{forcefactoor}
\end{eqnarray}
These terms describe the influence of the gravitational field on the particle
without taking into account its internal structure.
The first term in Eq. (\ref{forcefactoor}) is important for the motion
of the particle in nonstationary spacetimes, for example, in cosmological context.
The next term describes the Newtonian force, the related relativistic
corrections, and the Coriolis-like force proportional to $\bm K$. The last term
also contributes to the relativistic corrections to the force acting in static
spacetimes that arise in addition to the velocity-independent Newtonian force.

All the terms of the first order in the Planck constant 
are proportional to the spin operators and therefore they collectively
represent the quantum-mechanical counterpart of the Mathisson force
(which is an analogue of the Stern-Gerlach force in electrodynamics).
This force is given by, recall the notation (\ref{mathfrakT}),
\begin{equation}
 F_{\widehat a}^{(1)} = {\frac \hbar8} \left\{p_b,\left\{\mathfrak{T}^c,
\partial_cW^b{}_{\widehat a}\right\}\right\} - {\frac \hbar4}\left\{
W^b{}_{\widehat a},\partial_b{\cal U}\right\}.\label{forcefactoun}
\end{equation}
In the next section, we will demonstrate the agreement between the
quantum-mechanical and the classical equations of particle dynamics.

Eqs. (\ref{forcefactoor}) and (\ref{forcefactoun}) perfectly reproduce all 
previously obtained quantum-mechanical results \cite{PRD,PRD2,OST,ostrong}. 
In order to illustrate this, let us find the force on the spinning particle 
in the metric \cite{HN} of an arbitrarily moving noninertial (accelerated 
and rotating) observer:
\begin{equation}
V = 1 + {\frac {{\bm a}\cdot{\bm r}}{c^2}},\qquad W^{\widehat
a}{}_b = \delta^a_b,\qquad K^a =-{\frac 1c}\,(\bm\omega\times\bm
r)^a.\label{VWni}
\end{equation}
The FW Hamiltonian for this metric was derived in \cite{ostrong}. It reads: 
\begin{eqnarray}
{\cal H}_{FW} &=& \frac\beta2\left\{\left(1+\frac{\bm a\cdot\bm r}{c^2}\right),
\sqrt{m^2c^4+c^2\bm p^2}\right\}-\bm\omega\cdot({\bm r}\times{\bm p})\nonumber\\
&& +\,\frac\hbar2\bm\Pi\cdot\frac{\bm a\times\bm p}{mc^2(\gamma+1)}
-\frac\hbar2\bm\Sigma\cdot\bm\omega,\label{Hamltni}
\end{eqnarray}
where the object that has the meaning of the Lorentz factor is defined by
\begin{equation}
\gamma = \frac{\sqrt{m^2c^4+c^2\bm p^2}}{mc^2}.\label{gam}
\end{equation}
Using the FW Hamiltonian (\ref{Hamltni}) in (\ref{forcefacto}) and (\ref{dpH})
yields the explicit force 
\begin{eqnarray}
F_{\widehat a} &=& -\,{\frac \beta{c^2}}\,a_a\sqrt{m^2c^4+c^2\bm p^2} 
- ({\bm\omega}\times{\bm p})_a\nonumber\\
&=& \beta m\gamma\left(-\,\bm{a} + \bm{v}\times\bm{\omega}\right)_a.\label{MP4}
\end{eqnarray}
Here we used (\ref{gam}) and the relation between the operators of momentum 
and velocity $p_{\widehat a} = e_{\widehat a}^bp_b = \beta\gamma m v_a$ which is 
recovered from (\ref{Fpv}). One can straightforwardly verify that the usual
structure of the inertial forces (in particular, the Coriolis and centrifugal 
pieces) is encoded in the force (\ref{MP4}), see the corresponding computation
of the coordinate acceleration operator in \cite{ostrong}. 

For the metric (\ref{VWni}), the spacetime curvature vanishes. As a result,
the curvature- and spin-dependent Mathisson force is zero. In the general case, 
the Mathisson force is nontrivial, and the validity of the equivalence principle 
is an open question (see, e.g., Ref. \cite{Plyatsko:1997gs}). In a separate 
publication, we will analyse the possible generalization of the equivalence 
principle for spinning particles, making use of the force framework developed 
here. As a preliminary step, we refer to \cite{ostrong} where we evaluated the 
quantum force for the weak gravitational field and recovered the linearized 
Mathisson force, thus confirming the earlier results \cite{wald,barker}.

Any theory based on the Dirac equation can reproduce only a
certain reduced form of the equation of spin motion. The formal reason is the
absence in the Lagrangian and the Hamiltonian of the terms bilinear in the spin
matrices because their product can always be simplified: $\Sigma^a\Sigma^b =
\delta^{ab} + i\epsilon^{abc}\Sigma^c$. As a result, the equation of spin motion
of a Dirac particle cannot contain such terms.
In quantum mechanics of particles with higher spins $(s>1/2)$ as well as in
the classical gravity, the terms bilinear in spin cannot be reduced and the
general MP equations \cite{Mathisson,Papapetrou} should be used.

\section{Classical spinning particles}\label{CS}

\subsection{Mathisson-Papapetrou approach}

The motion of classical spinning particles in the gravitational field
can be consistently described by the generally covariant MP theory
\cite{Mathisson,Papapetrou}, for the recent discussion see also
\cite{chicone,dinesh,plyatsko,plyatsko2}). In this framework, a test particle is
characterized by the 4-velocity $U^\alpha$ and the tensor of spin $S^{\alpha\beta}
= - S^{\beta\alpha}$. The total 4-momentum is not collinear with the velocity,
in general. In \cite{PK}, a different noncovariant approach was developed,
in which the main dynamical variable is the 3-dimensional spin defined in
the rest frame of a particle. In our previous work
\cite{PRD,PRD2,Warszawa,OST,ostrong} we have used the MP theory,
and demonstrated its consistency with the noncovariant approach.

The analysis of the general MP equations is a difficult task \cite{plyatsko}
and the exact solutions are not available even for the simple spacetime
geometries. The knowledge of the symmetries of the gravitational field,
i.e., of the corresponding Killing vectors, significantly helps in the
integration of the equations of motion, as can be demonstrated \cite{OP}
for the de Sitter spacetime, in particular. However, in the absence of
the symmetries, various approximation schemes were developed to find
solutions of MP equations of motion. Following \cite{chicone}, we
neglect the second order spin effects, so that the MP system reduces to
\begin{eqnarray}\label{dP}
{\frac {DU^\alpha}{d\tau}} &=& f_{\rm m}^\alpha = -\,{\frac 1{2m}}
\,S^{\mu\nu}U^\beta R_{\mu\nu\beta}{}^\alpha,\\
{\frac {DS^{\alpha\beta}}{d\tau}} &=& {\frac {U^\alpha U_\gamma}{c^2}}
{\frac {DS^{\gamma\beta}}{d\tau}} -  {\frac {U^\beta U_\gamma}{c^2}}
{\frac {DS^{\gamma\alpha}}{d\tau}}.\label{dS}
\end{eqnarray}
On the right-hand side of (\ref{dP}) we have the Mathisson force
$f_{\rm m}^\alpha$ that depends on the Riemann curvature tensor
$R_{\mu\nu\beta}{}^\alpha$ of spacetime. The tensor of spin satisfies
the Frenkel condition $U_\alpha S^{\alpha\beta} = 0$ and gives rise
to the vector of spin
\begin{equation}\label{Sa}
S_\alpha = {\frac 12}\epsilon_{\alpha\beta\gamma}S^{\beta\gamma}.
\end{equation}
Here we use the totally antisymmetric tensor
\begin{equation}\label{epsabc}
\epsilon_{\alpha\beta\gamma} = {\frac 1c}\eta_{\alpha\beta\gamma\delta}U^\delta,
\end{equation}
constructed from the Levi-Civita tensor $\eta_{\alpha\beta\gamma\delta}$.
The relation (\ref{Sa}) can be inverted
\begin{equation}\label{Sab}
S^{\alpha\beta} = - \,\epsilon^{\alpha\beta\gamma}S_{\gamma}
\end{equation}
with the help of the identity
\begin{equation}
\epsilon^{\alpha\beta\gamma}\epsilon_{\mu\nu\gamma} = P^\alpha_\nu P^\beta_\mu
- P^\alpha_\mu P^\beta_\nu,\label{eePP}
\end{equation}
where $P^\alpha_\mu = \delta^\alpha_\mu - {\frac 1{c^2}}U^\alpha U_\mu$
is the projector on the rest frame of the particle.

Using the definition (\ref{Sa}), we rewrite the equation (\ref{dS})
in an alternative form
\begin{equation}\label{dSa}
{\frac {DS_{\alpha}}{d\tau}} = {\frac {U_\alpha U^\beta}{c^2}}
{\frac {DS_\beta}{d\tau}} = -\,{\frac 1{c^2}}U_\alpha f_{\rm m}^\beta S_\beta.
\end{equation}

With the help of the projectors and antisymmetric tensor, one can
decompose the curvature tensor into the three irreducible pieces
\begin{eqnarray}
\dE_{\alpha\beta} &=& {\frac {U^\mu U^\nu}{c^2}}R_{\alpha\mu\beta\nu},\label{Eab}\\
\dM^{\alpha\beta} &=& {\frac 14}\epsilon^{\alpha\mu\nu}\epsilon^{\beta\rho\sigma}
R_{\mu\nu\rho\sigma},\label{Mab}\\
\dB_\alpha{}^\beta &=& {\frac {U_\gamma}{2c}}\epsilon_{\alpha\mu\nu}
R^{\beta\gamma\mu\nu}.\label{Bab}
\end{eqnarray}
By construction, these tensors satisfy the orthogonality conditions
$\dE_{\alpha\beta}U^\beta = 0$, $\dM^{\alpha\beta}U_\beta = 0$, $\dB_\alpha{}^\beta
U_\beta = 0$, $\dB_\alpha{}^\beta U^\alpha = 0$. Taking into account the
obvious symmetry $\dE_{\alpha\beta} = \dE_{\beta\alpha}$ and $\dM^{\alpha\beta}
= \dM^{\beta\alpha}$, we have $6+6+8=20$ independent components for these
objects. The curvature decomposition reads explicitly
\begin{eqnarray}
R^{\alpha\beta\mu\nu} &=& {\frac 1{c^2}}\left(U^\alpha U^\mu \dE^{\beta\nu}
- U^\beta U^\mu \dE^{\alpha\nu} - U^\alpha U^\nu \dE^{\beta\mu} + U^\beta U^\nu
\dE^{\alpha\mu}\right) + \epsilon^{\alpha\beta\gamma}\epsilon^{\mu\nu\lambda}
\dM_{\gamma\lambda}\nonumber\\
&& +\,{\frac 1c}\left[\epsilon^{\alpha\beta\gamma}\left(U^\mu\dB_\gamma{}^\nu
- U^\nu\dB_\gamma{}^\mu\right) + \epsilon^{\mu\nu\gamma}\left(U^\alpha
\dB_\gamma{}^\beta - U^\beta\dB_\gamma{}^\alpha\right)\right].
\end{eqnarray}
As a result, we rewrite the Mathisson force as
\begin{equation}
f_{\rm m}^\alpha = {\frac c{2m}}\dB_\beta{}^\alpha S^\beta.\label{fM}
\end{equation}

The physical spin is defined in the rest frame of a particle where the
4-velocity reduces to $u^\alpha = (1, \bm{0}) = \delta^\alpha_0$. The local
reference frame and the rest frame are related by the Lorentz transformation
such that $U^\alpha = \Lambda^\alpha{}_\beta u^\beta$. Recalling $U^\alpha =
(\gamma, \gamma v^a)$, the Lorentz matrix reads explicitly
\begin{equation}\label{Lab}
\Lambda^\alpha{}_\beta = \left(\begin{array}{c|c}\gamma & \gamma v_b/c^2 \\
\hline \gamma v^a & \delta^a_b + (\gamma - 1)v^av_b/v^2\end{array}\right),
\end{equation}
with the Lorentz factor $\gamma = 1/\sqrt{1 - v^2/c^2}$, where $v^2 =
\delta_{ab}v^av^b$.

The physical spin is then $s^\alpha = (\Lambda^{-1})^\alpha{}_\beta S^\beta$, hence
$s^\alpha = (0, \bm{s})$. We rewrite equation (\ref{dSa}) as ${\frac {dS^\alpha}
{d\tau}} = \Phi^\alpha{}_\beta S^\beta$, with $\Phi^\alpha{}_\beta = - U^i
\Gamma_{i\beta}{}^\alpha + {\frac {1}{c^2}}(f_{\rm m}^\alpha U_\beta - f_{\beta{\rm m}}
U^\alpha)$. From this we find the equation of motion of the physical spin:
 \begin{eqnarray}
{\frac {ds^\alpha}{d\tau}} &=& \Omega^\alpha{}_\beta s^\beta,\label{dsp}\\
\Omega^\alpha{}_\beta &=& (\Lambda^{-1})^\alpha{}_\gamma\Phi^\gamma{}_\delta
\Lambda^\delta{}_\beta - (\Lambda^{-1})^\alpha{}_\gamma{\frac d {d\tau}}
\Lambda^\gamma{}_\beta.\label{Oprec}
\end{eqnarray}

Noticing that with respect to the coordinate basis the 4-velocity is
$U^i = \gamma e^i_{\widehat 0}+ \gamma v^a\,e^i_{\widehat a}$,
we recast the MP system (\ref{dP}) and (\ref{dsp}) into the 3-vector form
\begin{eqnarray}
{\frac {d\gamma}{d\tau}} &=& {\frac \gamma {c^2}}\,{\bm v}\cdot
\widehat{\bm{\mathcal{E}}},\label{dgamma}\\
{\frac {d(\gamma\bm{v})}{d\tau}} &=& \gamma\left(\widehat{\bm{\mathcal{E}}} 
+ \bm{v}\times\bm{\mathcal{B}}\right),\label{force}\\
{\frac {d{\bm s}}{d\tau}} &=& \bm\Omega\times\bm s.\label{dsMP}
\end{eqnarray}
Here using (\ref{LT}) and (\ref{fM}), we introduced the objects that can be 
called the generalized gravitoelectric and gravitomagnetic fields:
\begin{eqnarray}\label{ge}
{\mathcal E}^a &=& {\frac {\gamma}V}\delta^{ac}\left(c{\cal Q}_{(\widehat{c}\widehat{b})}
v^b - c^2\,W^b{}_{\widehat{c}}\,\partial_bV\right),\\ \label{gm}
{\mathcal B}^a &=& {\frac {\gamma}V}\left(-\,{\frac c2}\,{\Xi}^a -
{\frac 12}\Upsilon\,v^a + \epsilon^{abc}V{\cal C}_{bc}{}^dv_d\right),\\
\widehat{\bm{\mathcal{E}}}{}^a &=& \bm{\mathcal{E}}{}^a + {\frac {c}{2m\gamma}}
\dB_b{}^a\left(s^b - {\frac {\gamma}{\gamma + 1}}
{\frac {v^bv_c}{c^2}}s^c\right).\label{fMa}
\end{eqnarray}
The components of the angular velocity of the spin precession ${\bm\Omega}
= \left\{-\,{\frac 12}\epsilon^{abc}\Omega_{bc}\right\}$ are obtained from
(\ref{Oprec}):
\begin{equation}
\bm{\Omega} = -\bm{\mathcal{B}} + {\frac {\gamma}{\gamma +
1}}\,{\frac {\bm{v} \times\bm{\mathcal{E}}} {c^2}}.\label{omgem}
\end{equation}
Alternatively, we can explicitly write the precession velocity components
with the help of (\ref{connection1}) and (\ref{connection2}) as \cite{PK,OST} 
\begin{equation}
\Omega_{\widehat{a}} = \epsilon_{abc}\,U^i\left({\frac 12}\Gamma_i{}^{\widehat{c}
\widehat{b}} +{\frac {\gamma}{\gamma + 1}}\,\Gamma_{i\widehat{0}}{}^{\widehat{b}}
v^{\widehat{c}}/c^2\right).\label{OmegaG}
\end{equation} 
Finally, substituting (\ref{ge}) and (\ref{gm}) into (\ref{omgem}), 
we obtain the {\it exact classical formula} for the angular velocity of 
the spin precession in an arbitrary gravitational field:
\begin{eqnarray}
\Omega^{\widehat{a}} &=& {\frac {\gamma}{V}}\left({\frac 12}{\Upsilon}
\,v^{\widehat{a}} - \epsilon^{abc}V{\cal C}_{{\widehat{b}}{\widehat{c}}}{}^d
v_{\widehat{d}} + {\frac \gamma{\gamma + 1}}\epsilon^{abc}W^d{}_{\widehat{b}}
\,\partial_dVv_{\widehat{c}}\right.\nonumber\\
&& +\left.{\frac c2}\,{\Xi}^{\widehat{a}} - {\frac \gamma{\gamma +
1}} \epsilon^{abc}{\cal Q}_{(\widehat{b}\widehat{d})}{\frac
{v^{\widehat{d}} v_{\widehat{c}}} c}\right).\label{OmegaS}
\end{eqnarray}
The terms in the first line are linear in the 4-velocity of the
particle, whereas the terms in the second line contain the even
number of the velocity factors.

As compared to the  precession of the quantum spin described by
$\bm\Omega^{(1)}$ and $\bm\Omega^{(2)}$ using the coordinate time,
the classical spin precession velocity $\Omega^{\widehat{a}}$ contains
an extra factor 
\begin{equation}
{\frac {dt}{d\tau}} = U^0 = {\frac \gamma V},\label{dtdt}
\end{equation}
since the classical dynamics is parameterized using the proper time.

It is worthwhile to notice that the equations of motion of a particle
(\ref{dgamma}) and (\ref{force}) have a remarkably simple form of the
motion of a relativistic charged particle under the action of the
Lorentz force. It is interesting to mention a certain asymmetry: the
Mathisson force (\ref{fM}), that depends on the spin and the curvature
of spacetime, contributes only to the gravitoelectric field (\ref{fMa})
but not to the gravitomagnetic one. Using (\ref{dgamma}) in (\ref{force}),
we can recast the latter into the dynamical equation
\begin{equation}\label{MP}
{\frac {d\bm{v}}{d\tau}} = \widehat{\bm{\mathcal{E}}} - {\frac {\bm{v}(\bm{v}
\cdot\widehat{\bm{\mathcal{E}}})} {c^2}} + \bm{v}\times\bm{\mathcal{B}}. 
\end{equation}
Let us consider the motion of the classical particle in the metric of a 
noninertial observer (\ref{VWni}). Computing the gravitoelectric and 
gravitomagnetic fields is straightforward: $\widehat{\bm{\mathcal{E}}} = 
\bm{\mathcal E} = -\,{\frac \gamma V}\,\bm{a}$, and $\bm{\mathcal B} = 
{\frac \gamma V}\,\bm{\omega}$. As a result, 
\begin{equation}
{\frac {d(\gamma\bm{v})}{dt}} = \gamma\left(-\,\bm{a} + \bm{v}\times\bm{\omega}
\right),\label{force2}
\end{equation}
where we changed from the proper time parametrization to the coordinate time
using (\ref{dtdt}). As we see, the classical (\ref{force2}) and the quantum 
(\ref{MP4}) forces are the same. 

Finally, making use of (\ref{Fpv}) and (\ref{eV}),
we conclude that the classical equation of the spin motion (\ref{omgem})
agrees with the quantum equation (\ref{spinmeq}) and with the
semiclassical one (\ref{ds}). Thus, the classical and the quantum
theories of the spin motion in gravity are in complete agreement.
This is now verified for the {\it arbitrary gravitational field}
configurations. We thus confirm and extend our previous results
obtained for the weak fields \cite{OST} and for special strong field
configurations \cite{ostrong}.

\subsection{Hamiltonian approach}\label{canonical}

It is instructive to compare the classical and quantum Hamiltonians
of a spinning particle. In order to do this, one can start from the
classical Hamiltonian of a spinless relativistic point particle (with
an electric charge $q$, in general). The action has the well-known form
\begin{eqnarray}
I = - \int mc^2d\tau + qA_idx^i = -\int \left[mc\left(g_{ij}U^iU^j
\right)^{1/2} + qA_iU^i\right]d\tau.\label{I0}
\end{eqnarray}
In order to avoid working with the constrained system, we will use
the deparatmetrized formulation. With the 3-velocity $v^a = dx^a/dt =
U^a/U^0$, we then recast the action into $I = \int{\cal L} dt$, where
\begin{eqnarray}
{\cal L} = -mc\left(g_{00} + 2g_{0a}v^a + g_{ab}v^av^b\right)^{1/2}
- qA_0 - qA_bv^b.\label{L0}
\end{eqnarray}
The canonical momentum is
\begin{eqnarray}
p_a = {\frac {\partial {\cal L}}{\partial v^a}} = -\,{\frac {mc(g_{0a} + g_{ab}v^b)}
{(g_{00} + 2g_{0a}v^a + g_{ab}v^av^b)^{1/2}}} - qA_a.\label{pa}
\end{eqnarray}
Inverting, we find velocity in terms of momentum $\pi_a = p_a + qA_a$
\begin{eqnarray}
v^a = {\frac {g^{0a}}{g^{00}}} - {\frac {\tilde{g}^{ab}\pi_b}
{[g^{00}(m^2c^2 - \tilde{g}^{ab}\pi_a\pi_b)]^{1/2}}},\qquad  \tilde{g}^{ab}
= g^{ab} - \frac{g^{0a}g^{0b}}{{g}^{00}}.\label{va}
\end{eqnarray}
As a result, the classical Hamiltonian reads (we fix some sign errors of
\cite{Cogn} here):
\begin{equation}
{\cal H}_{class} = p_av^a - {\cal L} = \left(\frac{m^2c^2 - \tilde{g}^{ab}\pi_a\pi_b}
{{g}^{00}}\right)^{1/2} + \frac{g^{0a}\pi_a}{{g}^{00}} + qA_0.\label{clCog}
\end{equation}
For the contravariant components of the general metric (\ref{LT}) we have
$g^{ij} = e^i_\alpha e^j_\beta g^{\alpha\beta} = {\frac 1{c^2}}e^i_{\hat{0}}
e^j_{\hat{0}} - e^i_{\hat{c}}e^j_{\hat{d}}\delta^{cd}$. Thus explicitly, using
(\ref{frame}):
\begin{eqnarray}
g^{00} = {\frac {1}{c^2V^2}},\qquad g^{0a} = {\frac {K^a}{cV^2}},\qquad
g^{ab} = {\frac {1}{V^2}}\left(K^aK^b - {\cal F}^a{}_c{\cal F}^b{}_d
\delta^{cd}\right).\label{contra}
\end{eqnarray}
As a result, the classical Hamiltonian (\ref{clCog}) reads:
\begin{equation}
{\cal H}_{class}=\sqrt{m^2c^4V^2 + c^2\delta^{cd}{\cal F}^a{}_c\,{\cal F}^b{}_d
\,\pi_a\,\pi_b\,} + c{\bm K}\cdot{\bm\pi} + q\Phi.\label{clasp0}
\end{equation}

Now, let us discuss a generalization of the Hamiltonian theory with spin
included. In order to take into account the spin correctly, in a Cosserat type
approach a material frame (of four linearly independent vectors) is attached
to a particle, thus modelling its internal rotational degrees of freedom.
We denote it $h^i_\alpha$.

Such a material frame does not coincide with the spacetime frame, $h^i_\alpha
\neq e^i_\alpha$. In particular, the zeroth leg is given by particle's 4-velocity
\begin{equation}
h^i_{\widehat 0} = U^i.\label{hU}
\end{equation}
Any two orthonormal frames are related by a Lorentz transformation,
$h^i_\alpha = e^i_\beta\Lambda^\beta{}_\alpha$. The condition (\ref{hU}) means
that the Lorentz matrix $\Lambda^\beta{}_\alpha$ brings one to a local
reference frame $U^\alpha = \Lambda^\alpha{}_\beta u^\beta$ in which the particle
is at rest, i.e., $u^\alpha = \delta^\alpha_{\widehat 0}$. This is straightforwardly
demonstrated: $U^i = e^i_\alpha U^\alpha = e^i_\alpha \Lambda^\alpha{}_\beta u^\beta
= h^i_\alpha u^\alpha = h^i_{\widehat 0}$. The corresponding Lorentz transformation
is explicitly given by (\ref{Lab}).

The standard way to take the dynamics of spin into account
\cite{Porto,Bar,Stein} is to amend the classical Hamiltonian by
the term ${\frac 12}S^{ij}\Omega_{ij}$ with
\begin{eqnarray}\label{Omij}
\Omega^i{}_j := h^i_\alpha {\frac D{d\tau}} h^\alpha_j = h^i_\alpha U^k\nabla_k h^\alpha_j
= h^i_\alpha U^k\left(\partial_k h^\alpha_j - \Gamma_{kj}{}^lh^\alpha_l\right).
\end{eqnarray}
Rewriting everything in terms of the objects in particle's rest frame, $S^{\alpha\beta}
= h^\alpha_ih^\beta_jS^{ij}$ and $\Omega^\alpha{}_\beta = h_i^\alpha h_\beta^j\Omega^i{}_j$,
we find
\begin{eqnarray}
{\frac 12}S^{ij}\Omega_{ij} = {\frac 12}S^{\alpha\beta}\Omega_{\alpha\beta} = \bm s\cdot\bm\Omega.
\end{eqnarray}
Here we recover the precession velocity vector (\ref{OmegaG}).

The resulting complete Hamiltonian has the structure that was proposed in the
framework of the general discussion in the Ref. \cite{PK}:
\begin{equation}
{\cal H}_{class}=\sqrt{m^2c^4V^2 + c^2\delta^{cd}{\cal
F}^a{}_c\,{\cal F}^b{}_d \,\pi_a\,\pi_b\,} + c{\bm K}\cdot{\bm
\pi} + q\Phi+ \bm s\cdot\bm\Omega.\label{clasnewadded}
\end{equation}
In the general case, $\bm\Omega$ should include both electromagnetic and
gravitational contributions.

The obvious similarity of quantum (\ref{Hamlt}) and classical (\ref{clasnewadded})
Hamiltonians is another demonstration of complete agreement of the quantum-mechanical
and classical equations of motion discussed in the previous subsection.
The consistency between the classical Hamiltonian dynamics and the quantum-mechanical
equations of particle dynamics derived in Sec.~\ref{forces} is also confirmed by
the computation of the force. Switching off the electromagnetic field, we find
the classical equation for the force  
\begin{eqnarray}\label{forceclass}
F_{\hat a}^{class} = p_b\dot{W}^b{}_{\hat a} + p_b\,{\frac {\partial{\cal H}_{class}}
{\partial p_c}}\,\partial_c{W}^b{}_{\hat a} - {W}^b{}_{\hat a}\partial_b{\cal H}_{class}.
\end{eqnarray}
As we see, the equations (\ref{forcefacto}) and (\ref{forceclass}) completely
agree. In particular, rewriting the spin-dependent part in Eq.(\ref{forcefacto})
in terms of the spin operator, $\bm s=\hbar\bm\Sigma/2$, shows the consistency of
the corresponding parts in the two equations. 

\section{Conclusions}\label{final}

This paper continues the study of the motion of the quantum and classical Dirac
fermion particles with spin $1/2$ on a curved spacetime. Generalizing our earlier
findings in \cite{PRD,PRD2,Warszawa,OST,ostrong} obtained for the weak fields and
for the special static and stationary field configurations, we now consider the case
of an absolutely arbitrary spacetime metric. The convenient parametrization in terms
of the functions $V(t,x^c)$, $K^a(t,x^c)$, and $W^{\widehat a}{}_b(t,x^c)$ provides a
unified description of all possible inertial and gravitational fields. We also
include the classical electromagnetic field for completeness. In this general
framework, we derive the Hermitian Dirac Hamiltonian (\ref{Hamilton1}). Starting
with this master equation, we apply the Foldy-Wouthuysen transformation \cite{PRA}
and construct the Hamiltonian (\ref{eqFW}) in the FW representation for an {\it
arbitrary spacetime geometry}. In this paper, we have confined ourselves to the
purely Riemannian case of Einstein's general relativity theory, possible generalization
to the non-Riemannian geometries will be analysed elsewhere. Making use of the FW
Hamiltonian, we derive the operator equations of motion. In particular, we study
the quantum-mechanical spin precession (\ref{spinmeq}) and its semiclassical
limit (\ref{ds}). One can apply these general results to compare the dynamics of
a spinning particle in the inertial and gravitational fields, thus revisiting the
validity of the equivalence principle \cite{MO}. The also derive the force operator
and analyse the quantum dynamics of the particle under its action in Sec.~\ref{forces}.
In the second part of the paper, we consider the motion of the classical particle
with spin. In the framework of the Mathisson-Papapetrou theory, we obtained the 
dynamical equations (\ref{dgamma}), (\ref{force}) and (\ref{MP}) which have a 
remarkably simple form of the motion of a relativistic particle under the action 
of the Lorentz force, with the Mathisson force included into the generalized 
gravitoelectric field (\ref{fMa}). We also derived the equation (\ref{OmegaS}) for 
the angular velocity of spin precession in the general gravitational field. It is 
satisfactory to see that our results further confirm the earlier conclusions 
\cite{OST,ostrong} and demonstrate that the classical spin dynamics is fully 
consistent with the semiclassical quantum dynamics of the Dirac fermion. Finally, 
the complete consistency of the quantum-mechanical and classical descriptions of 
spinning particles is also established using the Hamiltonian approach in 
Sec.~\ref{canonical}.

Among the important issues that remain still open, we would like to mention the need
to carefully analyse the derivation of covariant equations of motion in the Dirac 
and in the Foldy-Wouthuysen representations. The crucial point in this study is to 
understand the definition of the position and spin operators in these two 
representations, in particular making use of the previous work on this subject in 
\cite{GIzvANB,wong3,Cianfrani,NW,dVFor,Dixon,Suttorp,Dekerf,Rivasbook,Stepanov,LFCosta}.

\section*{Acknowledgements}

This work was supported in part by JINR,
the Belarusian Republican Foundation for Fundamental Research
(Grant No. $\Phi$12D-002) and the Russian Foundation for Basic
Research (Grant No. 11-02-01538).

\end{document}